\documentclass[aps,prl,twocolumn,showpacs,superscriptaddress,groupedaddress]{revtex4-1}
\usepackage{graphicx}
\usepackage{mathtools}
\usepackage{bbold}
\usepackage{amssymb}
\usepackage{amsmath}
\usepackage{latexsym}
\usepackage{ulem}
\usepackage{color}
\usepackage{dcolumn}
\usepackage{subfigure}
\usepackage[T1]{fontenc}
\usepackage[utf8]{inputenc}
\usepackage[english,french]{babel}
\usepackage[colorlinks=true,linkcolor=blue,citecolor=blue]{hyperref} %
\usepackage{bm}        % for math
\usepackage{amssymb}   % for math

\usepackage[capitalize]{cleveref}

\begin{document}

\title{Magnetic Couplings in Half-Filled Bipartite Graphs}

\author{G. Bouzerar}
\email[E-mail:]{georges.bouzerar@neel.cnrs.fr}
%\author{M. Thumin}
\affiliation{Université Grenoble Alpes, CNRS, Institut NEEL, F-38042 Grenoble, France}                    
\date{\today}
%\clearpage
\selectlanguage{english}
\begin{abstract}
We break from perturbative frameworks to establish a non-perturbative generalization of magnetic exchange rules in half-filled bipartite networks. 
While historical models remain strictly tied to the debatable perturbation approach in the dilute limit and require the absence of zero-energy modes, our universal mechanism governs exchange interactions across arbitrary bipartite graphs. By bypassing traditional expansion bottlenecks, this framework provides an analytical predictive tool for magnetism in complex lattices, offering key design principles to engineer stable, long-range order in next-generation 2D spintronic devices.
\end{abstract}
%\pacs{??}

\maketitle
{\it{Introduction-}} In recent years, there has been a surge of interest in systems exhibiting flat bands (FB) at or near the Fermi energy. This fascinating emerging field of research continues to attract considerable attention and has not yet been fully explored \cite{Leykam-2024,Bergholtz-2013,Rhim-2021,Derzhko-2015,Vicencio-2021,Regnault-2022,Hase-2023}. In FB systems the kinetic energy is entirely frozen, making electron-electron interactions the only relevant energy scale and thus paving the way for entirely novel quantum phases of matter. Notable examples of exotic physics include unconventional superconductivity of a quantum geometric nature, ferromagnetism, topological states and the fractional quantum Hall effect \cite{Cao-2018,Balents-2020,Peri-2021,Lin-2018,Neupert-2011,Peotta-2015,B-Thumin-2025, Mukherjee-2015}.
Bipartite lattices such as the Lieb lattice (CuO$_2$ plane in cuprates), the "dice" lattice, the holey graphene, or even the twisted bilayer graphene are particularly relevant candidates for studying the physics associated with flat bands. Indeed, they naturally exhibit such bands at $E = 0$ when the number of sites differs between the two coupled sublattices.
In bipartite graphs, chiral symmetry (sublattice symmetry) plays an important role and leads to interesting properties of the spectrum and the associated eigenstates.
In these bipartite systems featuring flat bands, half-filling of the flat bands plays a crucial role in establishing numerous key exact results.
For the single-band Hubbard model at half-filling (arbitrary repulsive interaction amplitude) on a bipartite lattice, Lieb established that the ground state exhibits spontaneous ferrimagnetic order and has a total spin
\(S = \vert{}N_A - N_B\vert{}/2\) \cite{Lieb-1989}.
Furthermore, Mielke \cite{Mielke-1991} and Tasaki \cite{Tasaki-1992} have also rigorously demonstrated that if the FB has the lowest energy (or the highest, in the case of hole doping) and is half-filled, then regardless of the strength of the repulsive interaction, the ground state is ferromagnetic and saturated.
Recently, several important results concerning flat-band superconductivity have been demonstrated for the case of an attractive interaction in half-filled systems; these results remain valid even in the presence of disorder that preserves the bipartite character of the lattice.
\\
\begin{figure}[h!]\centerline
{\includegraphics[width=0.7\columnwidth,angle=0]{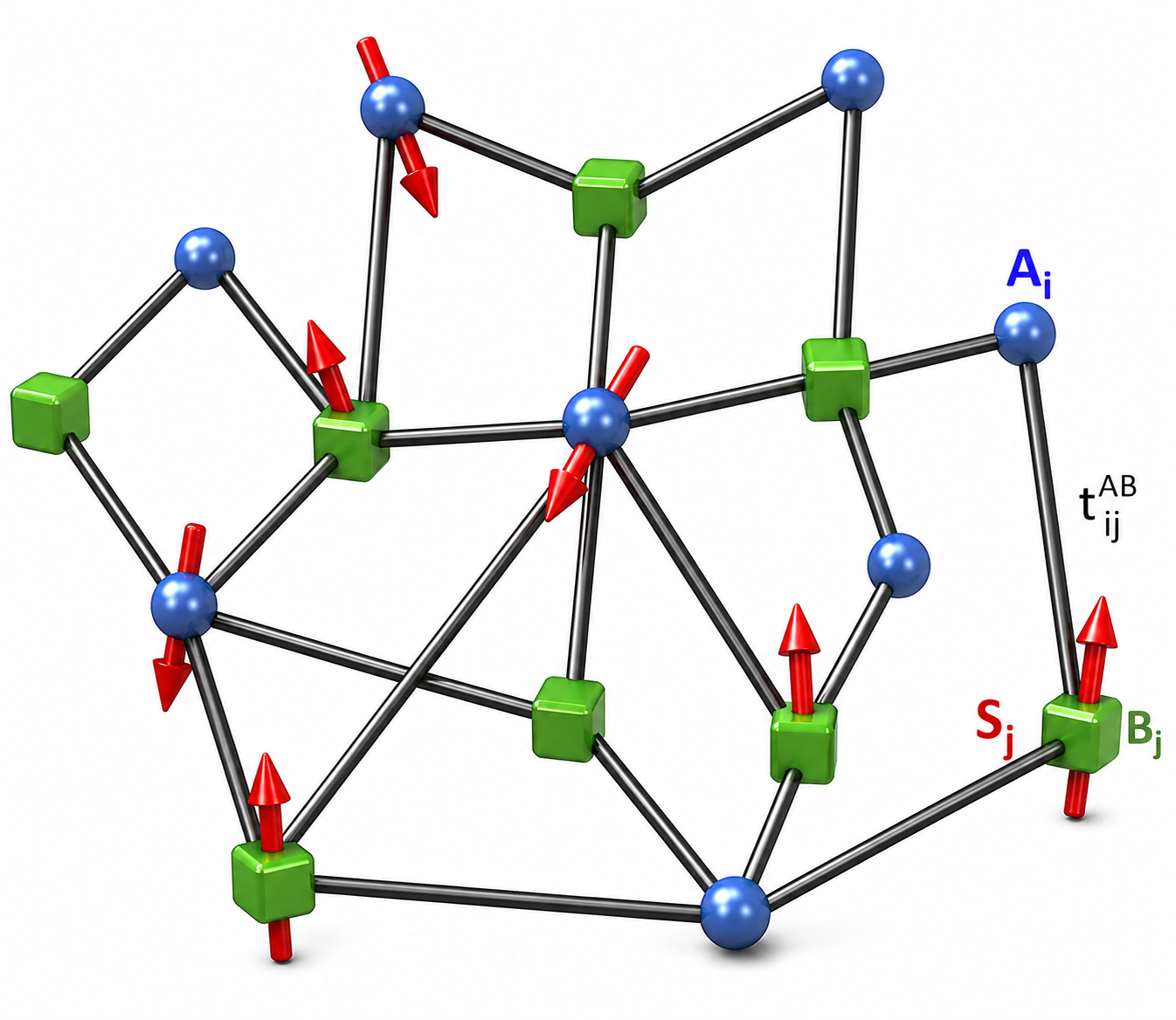}}
\vspace{-0.1cm} 
\caption{ Illustration of a magnetic bipartite graph. The green (respectively blue) symbols correspond to $B_j$ with $j=1,..,\Lambda_B$ (respectively $A_i$ with $i=1,..,\Lambda_A$) and the dark lines to the hoppings $t^{AB}_{ij}$.
The red arrows are the localized spins. 
}
\label{fig1}
\end{figure}

The nature of magnetic couplings also exhibits distinctive features in half-filled bipartite networks. It is well known that, in metallic systems, regardless of the degree of filling, a non-bipartite lattice exhibits oscillatory behavior known as “RKKY oscillations,” characterized (in a simple model) by $k_F$, the wave vector at the Fermi energy. If the lattice is bipartite and half-filled, the situation is radically different.
Indeed, within a second order perturbation theory it has been shown by Saremi \cite{Saremi-2007} that the coupling between pairs of localized spins is ferromagnetic when the spins are located on the same lattice and antiferromagnetic otherwise. 
However, as discussed in references \cite{Bunder-Lin-2009} and \cite{Lu-Lin-2014}, a problem arises in the perturbative calculations when energy modes at $E = 0$ are present in the host material's band structure. The rules governing the nature of the couplings break down.
It has even been demonstrated that FBs can induce strong frustration effects \cite{GB-2021}, leading to the destruction of the ferromagnetic order.
More recently \cite{GB-2023,Laubscher-2023}, it was shown that the major difficulty lies in standard second-order perturbation theory. Indeed, in the presence of FBs at the Fermi level, perturbation theory breaks down completely. However, a non-perturbative approach makes it possible to restore the sign rule for magnetic couplings, thanks in particular to crucial terms (FB-FB contributions) that are absent from the perturbative calculation. 
It should be noted, however, that reference \cite{GB-2023} focuses primarily on the case of the (two dimensional) Lieb lattice, whereas reference \cite{Laubscher-2023} focuses on specific one-dimensional FB systems, but no general proof has been established to date. The aim of this article is to provide a general proof established within a non-perturbative framework, valid for any bipartite graph, and any concentration of magnetic impurities and circumventing the problem associated with the presence of zero-energy modes.

{\it{The magnetic bipartite graph and its Hamiltonian-}}
The magnetic bipartite graph (MBG) as illustrated in Fig.\ref{fig1} consists in two sublattices $\cal{A}$ and $\cal{B}$, which contain respectively $\Lambda_A$ and $\Lambda_B$ orbitals.
We emphasize that the MBPG is not  restricted to ordered lattices. 
These two sets of orbitals are denoted $\cal{A}$ = $\{A_1, A_2,..., A_{\Lambda_A}\}$ and $\cal{B}$ = $\{B_1, B_2,..., B_{\Lambda_B}\}$.
$N_{imp}$ ($N_{imp} \le \Lambda =\Lambda_A+\Lambda_B$) localized magnetic moments of amplitude $S$ are considered located on both $\cal{A}$ and $\cal{B}$.

The Hamiltonian that describes itinerant carriers interacting with the localized spins reads, 
\begin{eqnarray}
\hat{H}=\sum_{i,j,\sigma}t^{AB}_{ij}c_{A_i,\sigma}^{\dagger}c^{}_{B_j,\sigma} + J\sum_{l\in \cal{I}_A} \hat{\bf s}^{A}_{l}\cdot{\bf{S}}^{A}_{l}
\label{hamilt}
\\ \nonumber
+ J\hspace{-0.0cm}\sum_{p\in \cal{I}_B} \hat{\bf s}^{B}_{p}\cdot{\bf{S}}^{B}_{p}
\end{eqnarray}
Because of the bipartite nature of the lattice, the only non vanishing hoppings are of the form $t^{AB}_{ij}$. For simplicity, this has no impact on our conclusions, we consider real hoppings. The operator $c_{\lambda_l,\sigma}^{\dagger}$ creates an electron of spin $\sigma$, in the orbital $\lambda_l$. The last two terms of the Hamiltonian describe the coupling between the localized classical spins ${\bf S}^{A}_l$ (respectively ${\bf S}^{B}_p$) at site $A_l$ (respectively $B_p$) and the quantum spin ($\hat{\bf s}^{A}_{l}$ or $\hat{\bf s}^{B}_{p}$) of the itinerant carrier. Finally, $\cal{I}_A$ (respectively $\cal{I}_B$) denotes the subset of sites of $\cal{A}$ (respectively $\cal{B}$) where localized spins are located.

In this study, we focus our attention on the case of half-filled systems for which the total number of electrons is $ \Lambda$. We also assume that the magnetic spin texture of ground state is ferrimagnetic, which implies that ${\bf S}^{A}_l = +S \textbf{e}_z$ ($l\in\cal{I}_A$) and 
${\bf S}^{B}_p = -S \textbf{e}_z$ ($p\in \cal{I}_B$).
This assumption is physically sound for bipartite lattices, aligning perfectly with the exact ground-state properties of the repulsive Hubbard model proven in refs. \cite{Lieb-1989, Shen-1994}.
Rigorously demonstrating, within our model and within a general framework, that this spin texture minimizes the ground state energy of the half filled system is a non trivial task and represents a genuinely interesting challenge.
However, as shown in Ref. \cite{GB-2023}, even if the initially chosen spin texture does not correspond to the true magnetic ground state, the calculation of the magnetic couplings will provide information on this matter. In what follows, the ferrimagnetic spin configuration is denoted 'AF'.
\\
For the AF spin texture, the Hamiltonian given in Eq.\eqref{hamilt} is rewritten,
$\hat{H}=\hat{H}^{\uparrow} + \hat{H}^{\downarrow}$, where,
\begin{equation} 
\hat{H}^{\sigma} = 
   \begin{bmatrix} 
         \hat{V}_{\sigma}^A & \hat{H}_{AB} \\ 
         \hat{H}_{AB}^{\dagger}& -\hat{V}_{\sigma}^B \\
   \end{bmatrix}.
   \label{hsigma}
\end{equation}
$\hat{H}_{AB}$ is the tight-binding term in Eq.$\,$\eqref{hamilt}. The spin dependent on-site potential matrices are $\hat{V}_{\sigma}^\lambda=\frac{1}{2}z_{\sigma}JS\,\text{diag}({\epsilon}_{\lambda1},{\epsilon}_{\lambda2},...,{\epsilon}_{\lambda{\Lambda_\lambda}})$, where $\lambda = A$ or $B$, and $\epsilon_{\lambda j} = 1$ 
if a localized spin is located at the impurity site, otherwise $\epsilon_{\lambda j}=0$.
In addition, we have introduced the variable $z_\sigma$ which is $1$ (respectively $-1$) for $\sigma = \uparrow$ (respectively $\sigma = \downarrow$). 

{\it{The nature of the zero energy modes-}}
Given that the presence of zero-energy modes (ZEMs) breaks the validity of Saremi’s formulation, we introduce foundational results to rigorously address this issue. First, in the uncoupled limit (\(JS = 0\)), the system can support ZEMs whose degeneracy is not inherently bounded by \(\vert\Lambda_B-\Lambda_A\vert\).
When present, these ZEM are strictly localized, having support exclusively on either sublattice \(\mathcal{A}\) or sublattice \(\mathcal{B}\). Let 
 $|\Psi_{0}\rangle = (|{\mathbf{a}_0}\rangle,|{\mathbf{b}_0}\rangle)^t$
an eigenstate of $H_\uparrow$ or $H_\uparrow$ ($|{\mathbf{a}_0}\rangle \in \cal{A}$ and $|{\mathbf{b}_0}\rangle \in \cal{B}$ ),
then by chiral symmetry the state $|\Psi'_{0}\rangle = 
\hat{\Gamma}|\Psi_{0}\rangle=
(|{\mathbf{a}}_0\rangle,|{-\mathbf{b}}_0\rangle)^t$ is as well a ZEM, where the block diagonal matrix
$\hat{\Gamma}= \begin{bmatrix}   \hat{\mathbb{1}}_{\Lambda_A}  & 0 \\ 0 & -\hat{\mathbb{1}}_{\Lambda_B}\ \end{bmatrix}$.
Thus, the linear combinations \(\frac{1}{\sqrt{2}}(\vert{}\Psi_{0}\rangle \pm \vert{}\Psi'_{0}\rangle)\), yields ZEMs whose support is exclusively restricted to sublattice \(\mathcal{A}\) or \(\mathcal{B}\), respectively.

\begin{figure}[h!]\centerline
{\includegraphics[width=1.1\columnwidth,angle=0]{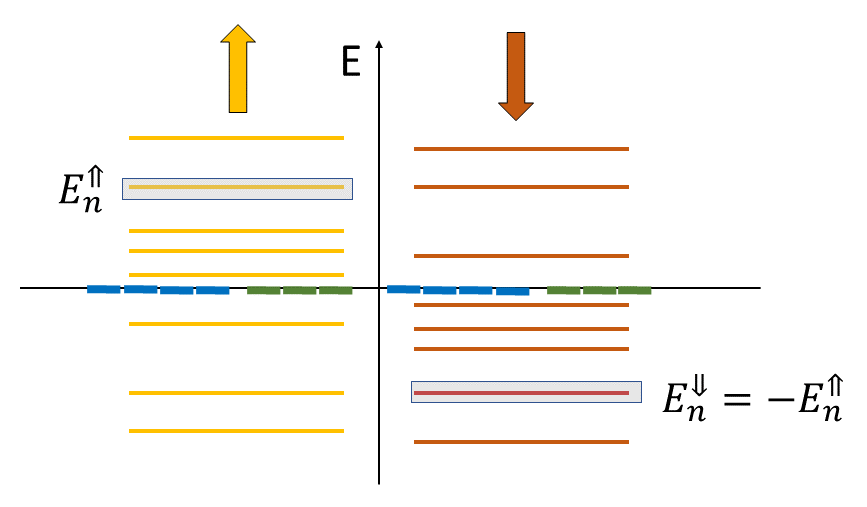}}
\vspace{-0.0cm} 
\caption{The spectrum of $\hat{H}^{\uparrow}$ and $\hat{H}^{\downarrow}$. The ZEMs are depicted in blue (respectively green) for those having a support on sublattice $\cal{A}$ only (respectively on $\cal{B}$ only). 
The ZEMs have zero weight at the sites occupied by magnetic impurities.
To each eigenstate of energy $E^{\uparrow}_n$ in $\uparrow$ sector corresponds an eigenstate of energy $E^{\downarrow}_n=-E^{\uparrow}_n$ in the $\uparrow$ sector.
}
\label{fig2}
\end{figure}
In the presence of a finite coupling between the carriers and the localized spins (\(JS \ne 0\)), the breaking of chiral symmetry significantly alters the zero-energy sector, causing some modes to lift while others reorganize. Crucially, we demonstrate that the remaining ZEMs exhibit a fundamental property: their wave-function weight systematically vanishes at all magnetic impurity sites.
Let $|\Psi_{0\uparrow}\rangle = (|{\mathbf{a}}_0\rangle,|{\mathbf{b}}_0\rangle)^t$ be
an eigenstate of $H_\uparrow$. This straightforwardly implies that,
\begin{equation} 
\hat{V}_{\uparrow}^A|{\mathbf{a}}_0\rangle =-\hat{H}_{AB}|{\mathbf{b}}_0\rangle
\label{eq1}
\end{equation}
and,
\begin{equation} 
\hat{H}_{AB}^{\dagger}|{\mathbf{a}}_0\rangle =\hat{V}_{\uparrow}^B|{\mathbf{b}}_0\rangle
\label{eq2}
\end{equation}
From Eq.~\eqref{eq1} and  Eq.~\eqref{eq2}, we immediately get $\langle {\mathbf{b}}_0 |\hat{V}_{\uparrow}^B|{\mathbf{b}}_0\rangle = -\langle {\mathbf{a}}_0 |\hat{V}_{\uparrow}^A|{\mathbf{a}}_0\rangle $, or equivalently, 
\begin{equation} 
\sum_{j\in \cal{I}_A} |\langle A_j|{\mathbf{a}}_0\rangle|^2 = -\sum_{l\in \cal{I}_B} |\langle B_l|{\mathbf{b}}_0\rangle|^2 
\label{eq3}
\end{equation}
Hence, the weight of the ZEM on the magnetic impurity sites is necessarily zero. Since Eq.~\eqref{eq3} is equivalent to $\hat{V}_{\uparrow}^A|{\mathbf{a_0}}\rangle=\hat{V}_{\uparrow}^B|{\mathbf{b_0}}\rangle=0$, then, the state $|\Psi'_{0\uparrow}\rangle = 
\hat{\Gamma}|\Psi_{0\uparrow}\rangle=
(|{\mathbf{a_0}}\rangle,-|{\mathbf{b_0}}\rangle)^t$ is as well a ZEM. Hence, even when $JS \ne 0$, the ZEM can be divided into two families: those located on $\cal{A}$ only and those on $\cal{B}$ excluding all the impurity sites.
Finally, we point out that the eigenstates in $\downarrow$ sector can be straightforwardly obtained from those
in the $\uparrow$ sector.
Indeed, if $|\Psi_{n\uparrow}\rangle = (|{\mathbf{a}_n}\rangle,|{\mathbf{b}_n}\rangle)^t$ is an eigenstate of energy $E_{n\uparrow}$ of $H_\uparrow$, then
$|\Psi_{n\downarrow}\rangle = \hat{\Gamma}|\Psi_{n\uparrow}\rangle=
(|{\mathbf{a}_n}\rangle,-|{\mathbf{b}_n}\rangle)^t$ is an eigenstate of energy $E_{n\downarrow} = -E_{n\uparrow}$ of $H_\downarrow$. The spectrum in both spin sectors is illustrated in Fig.\ref{fig2}.

{\it{The magnetic couplings-}}

To calculate the magnetic coupling (non-perturbative expression) between two spins located at $\textbf{R}_{i\lambda}$ and $\textbf{R}_{j\lambda'}$, we rely on the magnetic force theorem (MFT)~\cite{MFT1,MFT2}, widely used by the ab initio community and successfully implemented in a wide family of magnetic materials including disordered systems such as dilute magnetic semiconductors \cite{Sato-2010}. The MFT leads to,
\begin{eqnarray}
J_{\lambda\lambda'}(\textbf{R})=\frac{(JS)^{2}}{2}\int_{-\infty}^{+\infty} \chi_{ij}^{\lambda\lambda'}(\omega) f(\omega) d\omega, 
\label{eqcjij}
\end{eqnarray}
where $\textbf{R}=\textbf{R}_{i\lambda}-\textbf{R}_{j\lambda'}$ ($\lambda$ and $\lambda'$ being A or B) and the generalized susceptibility $\chi_{ij}^{\lambda\lambda'}(\omega)=-\frac{1}{\pi}
\operatorname{\Im}\left[G_{ij\uparrow}^{\lambda\lambda'}(\omega) G_{ji\downarrow}^{\lambda'\lambda}(\omega) \right]$. 
The Green's function $\widehat{G}_{\sigma}(\omega)=(\omega+i\eta-\widehat{H}^{\sigma})^{-1}$, $\eta$ mimics an infinitesimal inelastic scattering rate and $f(\omega)= \dfrac{1}{e^{(\omega-\mu)/k_BT}+1}$ is the Fermi-Dirac distribution. Here, we consider half-filled systems ($\mu=0$) at $T=0~K$.
It should be noted that the definition of the magnetic couplings given in Eq.\eqref{eqcjij} implies that ferromagnetic (respectively antiferromagnetic) exchange correspond to negative (positive) values. 
Furthermore, despite the explicit \((JS)^2\) factor in Eq. \eqref{eqcjij}, the exchange coupling does not inherently scale quadratically in the small-\(JS\) regime \cite{GB-2023}.
This equation can be rewritten,
\begin{eqnarray}
J_{\lambda\lambda'}(\textbf{R})=\frac{(JS)^{2}}{2}
\sum_{nl} C^{nl}_{i\lambda,j\lambda'} \frac{f(E^{\uparrow}_{n})-f(E^{\downarrow}_{l})}{E^{\uparrow}_{n}-E^{\downarrow}_{l}},
\label{eqcjijbis}
\end{eqnarray}
where the matrix element $C^{nl}_{i\lambda,j\lambda'} =
\langle i\lambda | \psi_{n\uparrow}\rangle
\langle \psi_{n\uparrow}| j\lambda'\rangle
\langle j\lambda'| \psi_{l\downarrow}\rangle
\langle \psi_{l\downarrow}| i\lambda \rangle$.
It is important to note that, since ZEM modes have zero weight at impurity sites, the matrix elements $C^{nl}_{i\lambda,j\lambda'}=0$ when $E^{\uparrow}_{n} = 0$ or $E^{\downarrow}_{l} = 0$. We can therefore completely ignore the ZEM modes, and Eq.~ $\eqref{eqcjijbis}$ remains well-defined. This equation can be divided into two different contributions $J_{\lambda\lambda'}(\textbf{R})=-\frac{1}{2}(JS)^{2}([1] + [2])$
where,
\begin{eqnarray}
[1] = \sum_{E^{\uparrow}_{n}>0,E^{\downarrow}_{l}<0}  \frac{C^{nl}_{i\lambda,j\lambda'}}{E^{\uparrow}_{n}-E^{\downarrow}_{l}} \\ \nonumber
[2] = \sum_{E^{\uparrow}_{n}<0,E^{\downarrow}_{l}>0}  \frac{C^{nl}_{i\lambda,j\lambda'}}{E^{\downarrow}_{l}-E^{\uparrow}_{n}}  
\label{eqcjijbis2}
\end{eqnarray}
Let us first focus on the first term and replace the eigenvalues and eigenvectors of the $\downarrow$-sector by those of the $\uparrow$-sector, using the symmetry pointed out in the previous section.
We can write, $[1] =\sum_{E^{\uparrow}_{n}>0,E^{\uparrow}_{m}>0}  \frac{C^{nm}_{i\lambda,j\lambda'}}{E^{\uparrow}_{n}+E^{\uparrow}_{m}}$
where the matrix element becomes,
$C^{nm}_{i\lambda,j\lambda'} =
\langle i\lambda | \psi_{n\uparrow}\rangle
\langle \psi_{n\uparrow}| j\lambda' \rangle
\langle j\lambda'| \psi_{m\uparrow}\rangle
\langle \psi_{m\uparrow}| i\lambda \rangle \eta_\lambda\eta_{\lambda'}
$, with $\eta_\lambda= 1$ (resp $-1$) if $\lambda \in \cal{A}$ (resp. $\cal{B}$).\\
By replacing $\frac{1}{E^{\uparrow}_{n}+E^{\uparrow}_{m}}$ by $\int_{0}^{\infty} e^{-s(E^{\uparrow}_{n}+E^{\uparrow}_{m})}ds$, we can write,

\begin{eqnarray}
[1] = \eta_\lambda\eta_{\lambda'}
\int_{0}^{\infty} \Big|F^{\lambda\lambda'}_{ij\uparrow}(s) \Big|^2ds
\end{eqnarray}
where, $F^{\lambda\lambda'}_{ij\uparrow}(s)=\sum_{E^{\uparrow}_{n}>0}\langle i\lambda| \psi_{n\uparrow}\rangle
\langle \psi_{n\uparrow}| j\lambda' \rangle e^{-sE^{\uparrow}_{n}}$.
Similarly, we find for the second term of Eq.~\eqref{eqcjijbis2},
\begin{eqnarray}
[2] = \eta_\lambda\eta_{\lambda'}
\int_{0}^{\infty} \Big|F^{\lambda'\lambda}_{ji\downarrow}(s) \Big|^2 ds
\end{eqnarray}
where $F^{\lambda'\lambda}_{ji\downarrow}(s)=\sum_{E^{\downarrow}_{n}>0}\langle j\lambda' | \psi_{n\downarrow}\rangle
\langle \psi_{n\downarrow}| i\lambda \rangle e^{-sE^{\downarrow}_{n}}$.

Thus, we end up with the key result of this article,
\begin{eqnarray}
\eta_\lambda\eta_{\lambda'}J_{\lambda\lambda'}(\textbf{R})=-\frac{1}{2}(JS)^{2}\times \nonumber \\
\Big( 
\int_{0}^{\infty} \Big|F^{\lambda\lambda'}_{ij\uparrow}(s)\Big|^2 ds\ +
\int_{0}^{\infty} \Big| F^{\lambda'\lambda}_{ji\downarrow}(s)\Big|^2 ds
\Big)
\label{eq-finale}
\end{eqnarray}
This single equation captures the main contribution of this work.
Consequently, magnetic impurities residing on the same sublattice dictate strictly ferromagnetic couplings (negative values), whereas cross-sublattice interactions inevitably yield antiferromagnetic alignment (positive values). This elegant outcome perfectly matches our initial premise of a ferrimagnetic ground-state spin texture.
Crucially, these findings establish a universal framework valid across all coupling regimes and any bipartite topology. This represents a major breakthrough that generalizes the foundational proof by S. Saremi \cite{Saremi-2007}, which was strictly confined to a two-impurity system treated via second-order perturbation theory with the strict constraint of absence of zero-energy modes (ZEM).
Remarkably, when the system exhibits degenerate states or flat bands at \(E=0\), this framework is no longer restricted to half-filling. The exchange coupling in Eq. \eqref{eq-finale} holds for an arbitrary partial filling of the zero-energy states
To demonstrate the concrete applicability of this generalized theory, we model a disordered two-dimensional magnetic system in the “Supplementary Material” \cite{SM}.

In conclusion, we have established a constraint-free framework that fundamentally redefines magnetic coupling in half-filled bipartite graphs. By overcoming the stringent limitations of traditional perturbative approach in the dilute limit and the absence of zero-energy modes, our approach unveils a universal mechanism for magnetism in complex lattices. Beyond its theoretical significance, this general framework provides a predictive blueprint for engineering robust, long-range magnetic order in two-dimensional materials. These findings may pave the way for a rational, model-driven design of next-generation spintronic architectures, including high-performance spin valves, spin transistors, and magnetic tunnel junctions.

%\begin{acknowledgments}
%\end{acknowledgments}

\vspace{3cm}

\section{Supplementary material for "Magnetic Couplings in Half-Filled Bipartite Graphs"}

\subsection{Introduction} 

\begin{figure}[h!]\centerline
{\includegraphics[width=1.0\columnwidth,angle=0]{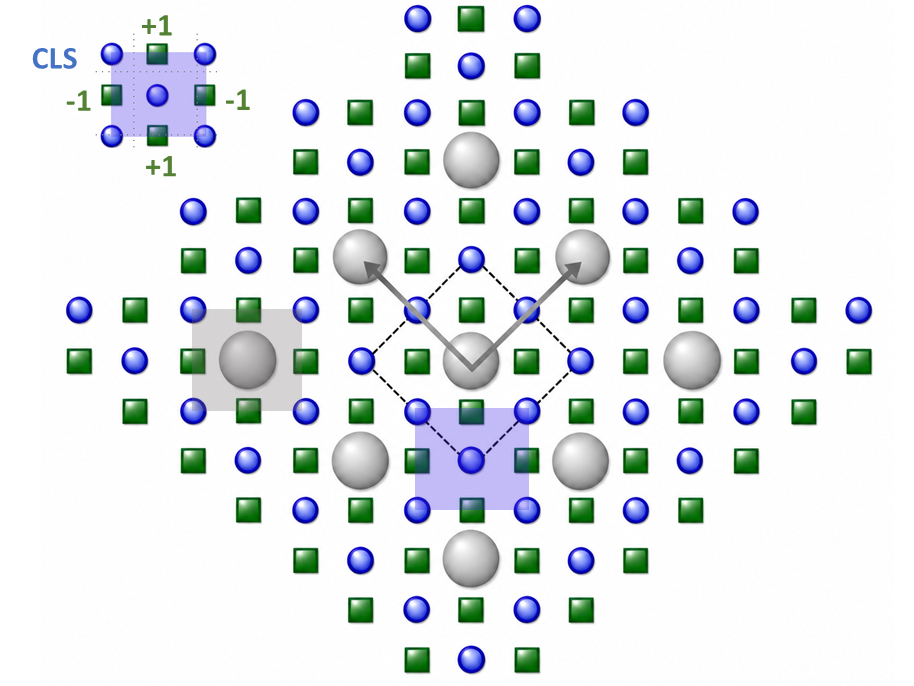}}
\vspace{-0.1cm} 
\caption{The depleted 1/8 square lattice. The unit cell (7 orbitals/cell) is indicated by the square with dashed lines and the removed sites of the square lattice (A-type) by the filled grey circles. Small filled blue circles correspond to orbitals of sublattice $\cal{A}$, and the filled green squares to orbitals of sublattice $\cal{B}$. 
The system contains two distinct plaquette configurations: gray squares centered on a vacancy and blue squares centered on a type-A atom. In the nearest-neighbor hopping case, the typical compact localized state (CLS) is localized within the latter configuration, sustaining a non-zero quantum amplitude solely on the four B orbitals of the blue plaquette.
}
\label{fig1-supp} 
\end{figure}

The purpose of this supplementary material is to provide an illustration of the main conclusions presented in the main text, focusing specifically on a two-dimensional disordered system that possesses zero energy modes.
To this end, we consider the case of the square lattice depleted by one-eighth, as illustrated in Fig.~\ref{fig1-supp}.
This lattice has seven orbitals per unit cell: 3 orbitals of type A and four of type B.
Without loss of generality we consider here that hoppings are restricted to nearest neighbours only, $t^{AB}_{ij} = t = -1$.  In the absence of magnetic impurities ($JS = 0$) the spectrum of this bipartite lattice consists in $6$ dispersive bands and a flat band located at $E = 0$. Compact localized states (CLS) associated with the flat band can be readily constructed within the plaquettes delimited by the vacancies as depicted in Fig.\ref{fig1-supp}. The CLS has non-zero support only on the B-type orbitals. 

In the second step, magnetic impurities are now distributed randomly across this lattice. More precisely, $N^{A}_{imp} = c_A \Lambda_{A}$ (respectively $N^{B}_{imp} = c_B \Lambda_{B}$) impurities are placed on sublattice $\cal{A}$ (respectively $\cal{B}$), which contain $\Lambda_A$ (respectively $\Lambda_B$) orbitals.
To distinguish between impurity sites belonging to $\cal{A}$ or $\cal{B}$, we introduce the variable $\eta_i$, which is equal to $1$ (or $-1$) if the impurity belongs to $\cal{A}$ (or $\cal{B}$). Finally, here, we have chosen the set of parameters $c_A = c_B = 0.3$ and $JS = 1$ (in unit of $t$). Note that this choice has no qualitative impact on the conclusions reached.

\subsection{The Hamiltonian} 

The real space Hamiltonian that describes the itinerant carriers interacting with the localized spins reads,
$\hat{H}=\hat{H}^{\uparrow} + \hat{H}^{\downarrow}$, where,
\begin{equation} 
\hat{H}^{\sigma} = 
   \begin{bmatrix} 
         \hat{V}_{\sigma}^A & \hat{H}_{AB} \\ 
         \hat{H}_{AB}^{\dagger}& -\hat{V}_{\sigma}^B \\
   \end{bmatrix}.
   \label{hsigmab}
\end{equation}
$\hat{H}_{AB}$ is the tight-binding term and the diagonal on-site potential matrices are $\hat{V}_{\sigma}^\lambda=\frac{1}{2}z_{\sigma}JS\,\text{diag}({\epsilon}_{\lambda_1},{\epsilon}_{\lambda_2},...,{\epsilon}_{\lambda{\Lambda_\lambda}})$, where $\lambda = A$ or $B$, and $\epsilon_{\lambda_j} = \pm1$ 
if a localized spin is located at the impurity site, otherwise $\epsilon_{\lambda_j}=0$, and finally  $z_\sigma =1$ (respectively -1) for $\sigma=\uparrow$ (respectively  $\sigma=\downarrow$).

As pointed out in the main text, we  recall that the eigenstates in $\downarrow$ sector can be straightforwardly obtained from those
in the $\uparrow$ sector.
Indeed, if $|\Psi_{n\uparrow}\rangle = (|{\mathbf{a}_n}\rangle,|{\mathbf{b}_n}\rangle)^t$ is an eigenstate of energy $E_{n\uparrow}$ of $H_\uparrow$, then
$|\Psi_{n\downarrow}\rangle = 
(|{\mathbf{a}_n}\rangle,-|{\mathbf{b}_n}\rangle)^t$ is an eigenstate of energy $E_{n\downarrow} = -E_{n\uparrow}$ of $H_\downarrow$. 
Hereafter, we define (i) the AF configuration as the one for which $\epsilon_{A_j} = 1$ for impurities located on $\cal{A}$ and $\epsilon_{B_l} = -1$ for those on $\cal{B}$ 
 and (ii) the F configuration as the one for which all spins are parallel.

\begin{figure}[h!]\centerline
{\includegraphics[width=1.0\columnwidth,angle=0]{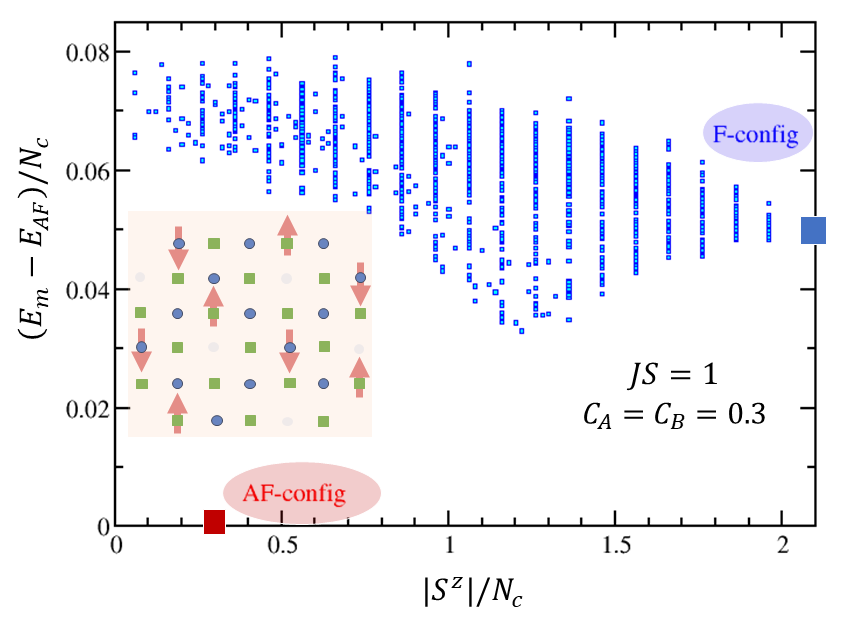}}
\vspace{-0.cm} 
\caption{For a given set of positions of the magnetic impurities, the energy difference (per cell), $(E_m - E_{AF})/N_c$ is plotted as a function of the total spin (per cell) of the magnetic configuration 'm'. The large blue square corresponds to the fully polarized spin texture (F) and the small blue squares correspond to random configurations. The values of $c_A$, $c_B$, and $JS$ are indicated in the figure.
}
\label{fig2-supp}
\end{figure} 

 \subsection{Ground-state energy and that of various spin texture} 
 
In Fig. \ref{fig2-supp}, the energy difference per cell, \((E_m - E_{AF})/N_c\), is plotted against the total spin per cell. Here, \(N_{c}\) signifies the total number of cells, \(E_{AF}\) is the energy  of the ferrimagnetic spin texture and
serves as the reference, \(E_{m}\) denotes the energies of alternative spin configurations.
First, one clearly sees that the AF spin texture minimizes the ground state energy of the system. As it can be seen, the energy of the F configuration (fully saturated ferromagnetic spin texture) is much larger than that of the AF configuration, here the difference of energy is about $\Delta E \approx 0.05~t$. We point out that these findings are qualitatively true for any values of the parameters $c_A$, $c_B$ and $JS$.

\subsection{Projected DOS and zero energy modes} 

\begin{figure}[h!]\centerline
{\includegraphics[width=1.0\columnwidth,angle=0]{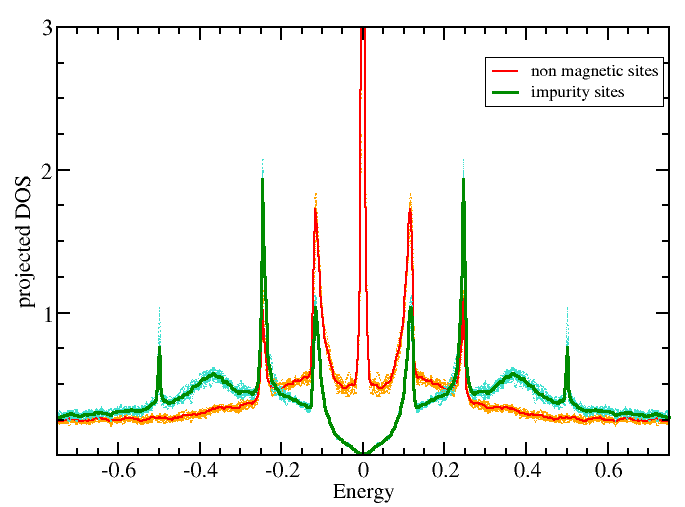}}
\vspace{-0.cm} 
\caption{Averaged Projected density of states (arbitrary unit) on impurity sites and on non magnetic sites as a function of the energy for the AF spin texture.
The orange dots and cyan ones correspond to different configuration of disorder (random position of the magnetic impurities). The red (projection on non magnetic sites) and green (projection on magnetic sites) continuous lines are the average over these set of configurations.
}
\label{fig3-supp}
\end{figure}

The average projected density of states (DOS), $\rho_{P}(E)=-\frac{1}{\pi N_c} \sum_{\sigma} Tr[ \hat{P}(E-H_\sigma)^{-1}\hat{P}]$, where $\hat{P}$ is the projector on the selected subset of orbitals is shown in Figure~\ref{fig3-supp} for the optimal antiferrimagnetic (AF) spin texture. We considered several random configurations for the positions of the magnetic impurities. More specifically, we performed the projection, on the one hand, over all sites containing magnetic impurities (type A or B) and, on the other hand, over the non-magnetic sites.
We observe a structure exhibiting multiple peaks. A well-defined peak at $E=0$ is clearly observed in the projected density of states (PDOS) on the non magnetic sites, indicating the presence of zero-energy modes (ZEMs) in the magnetic spectrum. On the other hand that projected on the non magnetic sites vanishes exactly at $E = 0$. This confirms that the ZEMs have no weight at the magnetic impurity sites. In other words, if $\vert \Psi_{0m}\rangle$ is a ZEM then at the impurity sites $\langle \lambda_i\vert \Psi_{0m}\rangle = 0$.
Notice that the peaks located exactly at $E = 0$ and $E = \pm JS/2$ can be well understood from the CLS depicted in Fig.~\ref{fig1-supp}. 
In fact, for non-zero \(JS\), the (blue) plaquettes lacking magnetic impurities at the B sites still correspond to eigenstates with energy \(E = 0\). In contrast, those blue plaquettes whose B sites are fully occupied by parallel-spin impurities support localized eigenstates with energy \(E = \pm JS/2\), one belongs to the $\uparrow$-sector and the other to the $\downarrow$-sector. 
As shown in Ref.~\cite{GB-2023}, for the (ordered) magnetic Lieb lattice, the contribution of these eigenstates to the couplings is crucial; it dominates in the small-\(JS\) regime, leading to a linear scaling of the couplings with \(\vert{}JS\vert{}\).

\subsection{Magnetic couplings in the half-filled disordered lattice}

We now propose to calculate the magnetic couplings using the expression given in Eq.(7) of the main text. The non perturbative expression of the magnetic coupling between two spins located at $r_{i\lambda}$ and $r_{j\lambda'}$, which relies on the magnetic force theorem ~\cite{MFT1,MFT2} is given by,
\begin{eqnarray}
J_{\lambda\lambda'}(\textbf{R})=\frac{(JS)^{2}}{2}\int_{-\infty}^{+\infty} \chi_{ij}^{\lambda\lambda'}(\omega) f(\omega) d\omega, 
\label{eqcjijb}
\end{eqnarray}
where, $\textbf{R}=\textbf{R}_{i\lambda}-\textbf{R}_{j\lambda'}$ ($\lambda$ and $\lambda'$ being A or B) and the generalized susceptibility $\chi_{ij}^{\lambda\lambda'}(\omega)=-\frac{1}{\pi}
\operatorname{\Im}\left[G_{ij\uparrow}^{\lambda\lambda'}(\omega) G_{ji\downarrow}^{\lambda'\lambda}(\omega) \right]$. 
The Green's function $\widehat{G}_{\sigma}(\omega)=(\omega+i\eta-\widehat{H}^{\sigma})^{-1}$, $\eta$ mimics an infinitesimal inelastic scattering rate and $f(\omega)= \dfrac{1}{e^{(\omega-\mu)/k_BT}+1}$ is the Fermi-Dirac distribution. Here, we consider half-filled systems ($\mu=0$) at $T=0~K$.
It should be noted  that the definition of the magnetic couplings given in Eq.~\eqref{eqcjij} implies that ferromagnetic (respectively antiferromagnetic) exchange correspond to negative (positive) values. 
\begin{figure}[h!]\centerline
{\includegraphics[width=1.0\columnwidth,angle=0]{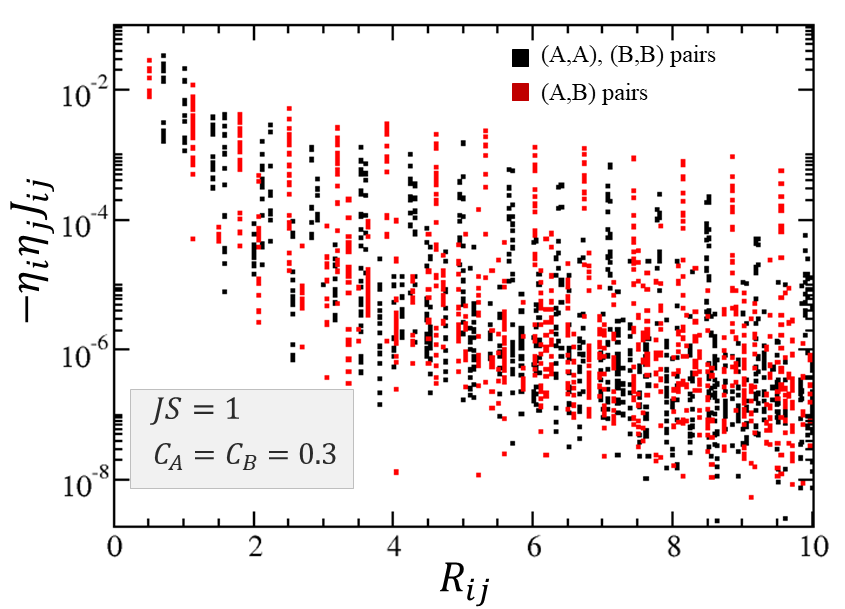}}
\caption{Couplings (in unit of t) in the disordered half-filled depleted square lattice as a function of the distance $R_{ij}$ between impurities. 
The coupling has been multiplied by the product $-\eta_i\eta_j$ (see text).
Red dots correspond to $(A,B)$ pairs and black ones to $(A,A)$ or $(B,B)$ pairs. We considered few random configurations for the positions of the magnetic impurities.
}
\label{fig4-supp}
\end{figure} 
The calculated couplings are depicted in Fig.~\ref{fig4-supp}. It can be clearly seen that the couplings are ferromagnetic when the localized spins are located on the same sublattice and antiferromagnetic otherwise. In any case $(A,A), (B,B)$ or $(A,B)$ pairs, we observe that the couplings decrease rapidly with distance, but they do not follow an exponential decay, as one might have expected given the absence of a band gap in the density of states. We wish as well to emphasize that, if the ferromagnetic spin texture is considered to be the ground state, it is also found that the magnetic couplings obey the same rules.
As previously mentioned, the ZEM density protects the exchange couplings shown in Fig. \ref{fig4-supp} against variation of the carrier concentration, making them perfectly robust throughout the range $\nu \in [\nu_{min}, \nu_{max}]$ around the half-filled case (\(\nu = 1\)). For the specific configuration considered here, we find $\nu_{min}= 0.94$ and 
$\nu_{max}= 1.04$.

\subsection{Conclusion}

In this supplementary material, we have highlighted—within the context of a disordered and depleted square magnetic lattice—the general rules governing the sign of magnetic couplings and the nature of ZEM states in half-filled bipartite lattices, which we have established within a general framework in the main text.

%\begin{acknowledgments}
%\end{acknowledgments}

\end{document}